\documentclass[12pt]{article}
\usepackage {amssymb}
\usepackage [dvips]{graphics}

\begin{document}
\baselineskip=15pt
\newtheorem{Corollary}{Corollary}

\begin{center}
{\bf Note on the typhoon eye trajectory}

{\it Olga Rozanova}\vskip1cm

Department of Differential Equations,
Mathematics and Mechanics Faculty     \\
Moscow State University                 \\
GSP-2 Vorobiovy Gory 119992 Moscow
Russia \\
E-mail: {\it rozanova@mech.math.msu.su} \vskip1cm {\bf Abstract}
\end{center}
{\it We consider a model of typhoon based on the three-dimensional
baroclinic compressible  equations of atmosphere dynamics averaged
over hight and describe a qualitative behavior of the vortex and
possible trajectories of the typhoon eye.}

\vskip2cm{ \bf 2000 Mathematics Subject Classification:}35Q30;
85A10; 76N15; 35L65.

\section{Preliminaries}

The typhoon (or tropical hurricane) is an intense vortex of middle
scale in the atmosphere having radius in several hundred
kilometers. It is a stable structure, existing sometimes within
more then two weeks. The typhoon trajectory can be very
complicated, it can suddenly change the direction, make loops, the
motion  stops sometimes for several days. The speed of wind in the
vortex region may amount to $100\, m/sec.$ The huge experimental
information, including the archive of hurricanes paths can be
found on the following sites: \cite{net}. The physical processes
inside of typhoon are complex \cite{Anthes},\cite{Hain}, for a
sufficient description of this phenomenon they use nonlinear
systems of PDE in three spatial dimensions, that can be solved
only numerically.

However, there are attempts to explain qualitative properties of
of  typhoon's behavior  by means of  simplified hydrodynamical
models. Note before all the outstanding paper by N.E.Kochin of
1922 \cite{Kochin}, where the cyclone model based on the
conditions of dynamic possibility in the sense of A.A.Fridman was
considered. The theory predicts that the trajectory may have loops
and the points of return. Earlier Rayleigh \cite {Rayleigh}, Show
\cite{Show}, Fridman\cite{Fridman} studied the trajectories of
atmospheric vortices.

There exists a trend where  methods of the solid state mechanics
are applied for a study of peculiarities of the tropical cyclones
trajectory. Here the tropical cyclone is modelled as a rigid
rotating cylinder (or a washer) in a flow of  viscous fluid and an
evaluation of different forces acting to the cylinder are made. In
\cite{Pur} the Coriolis force acting to the vertical component of
velocity inside of the typhoon is estimated, in \cite{Kuo, Shul,
Jones} the lifting (the Zhukovskii force), the friction against a
bedding surface and the drag force are considered, in \cite{Nest,
Khokh} all forces mentioned are taken into account. However, under
this approach the problem is a determination of the main flow
(generally speaking, it is assumed as given). If the cyclone
trajectory deviates from the trajectory of the flow, the Coriolis
force, lifting and drag forces begin to act, it may result the
appearance of oscillations and loops in trajectory in dependence
on parameters.

In a series of papers \cite{orda1, orda2, orda3, orda4} the method
of self-adapting model is applied to the study of typhoons
trajectories. The key point is the following: to determine
parameters  of a system of equations describing the motion of a
tropical typhoon, unknown in advance, they use the typhoon
trajectory known before a certain moment of time. As new
information becomes available, the values of parameters have been
adjusted, it gives a possibility to take into account their
time-dependence. The baric field is given as a polinomial in the
spatial variables with time-dependent coefficients, it is linear
in the simplest case.

In \cite{Dobryshman1, Dobryshman}, a non stationary axially
symmetric model of typhoon eye taking into account the vertical
processes is considered. However, only the three-dimensional
velocity takes part in the model, while the changing of density
and the Earth' rotation are not taken into account (whereas, they
are very essential factors, as we shall see below). The model
gives a possibility to find decaying and blowing up solutions (our
model predicts them as well).

Among relatively recent works it is necessary to mention a series
of papers \cite{BVDD,BVDD1,BVDD2,Dobrokhotov}. Here the "shallow
water" model was used (it coincides with the two-dimensional gas
dynamic equations, where the adiabatic exponent equals $2$). The
typhoon eye is considered as a point-singularity (week singularity
of square root type upon the Maslov hypothesis \cite{Maslov}). The
propagation of this singularity are defined with the necessity by
an infinite chain of ordinary differential equations. The
important problem is to find the method of closing the chain.
Authors use the approach proposed in \cite{11Ravi}. Namely,
according to this approach one can approximate the functions,
defining the solution and the singularity, by means of several
first terms of the Taylor series expansion near the origin of
coordinate system connected with the typhoon eye. The principal
point is that all this function are approximated by terms of same
order. After this procedure the system can be solved explicitly
for the first approximation (however, the predicted trajectory is
not realistic), and only numerically for the second approximation.
In the last case, by comparing the numerical data with the
trajectory of the real typhoon, it was observed their quite good
qualitative coincidence. Note that the ideas of self-adapting
model  can be used here to find the correct initial data for
computation \cite{BVDD}. Namely, given the trajectory of a real
typhoon, on can choose initial data such that the computed
trajectory is close by the real one. It is natural to assume that
these two trajectories will be similar within some next time.

In all theoretic work that we have mentioned phase transitions are
not taken into account, however its influence is important. In
this connection let us cite
 \cite{Moiseev}, where it is shown that the process of phase transitions
 result in an appearance of additional source of energy and a
 significant decreasing of the instability threshold.

Besides of hydrodynamic methods of the typhoon trajectory , there
exist statistical methods \cite{Gruza}, here we will not concern
them.

The present work is inspired by \cite{BVDD,BVDD1,BVDD2} and
correlates with these papers, however, it seems more plausible.
Let us clarify this.

The idea of consideration a typhoon as "a singularity of square
root type" is criticized by meteorologists. In \cite{Dobryshman}
it is noticed, that a deviation of pressure from the norm by
8-17\% in the center of typhoon is more likely an occasion for
linearization, not for " algebraic singularity of root type". Also
in \cite{Dobryshman} it is argued that modelling a typhoon one
cannot consider the horizontal scale greater, then vertical,
however, this assumption is necessary for the derivation of the
"shallow water" equations.

We use a two-dimensional model, which can be obtained from the
primitive three-dimensional system of atmosphere dynamics  taking
into account the Earth's rotation by means of averaging over hight
using the geostrophic approximation \cite{Alishaev}. Formally it
looks like the usual two-dimensional Navier-Stokes system, however
the "adiabatic exponent" is other (less, then the real one). Thus,
the vertical convective processes are "hidden", but taken into
account. We do not consider the process of the typhoon generation,
no doubt, it is principally three-dimensional, we suppose that the
vertex is formed and will exist within a certain time. Thus, it is
assumed implicitly that after the vertex formation the vertical
flows become well-balanced and allow the averaging. Our aim is to
analyze a possible displacement of the central domain of vortex
(the typhoon eye). We seek the exact solution of the system of
some special form. In fact, this signifies that we linearize the
velocity near the origin of the moving coordinate system
(experimental data give us such hint \cite{IAV}.)

The system of ODE, obtained in this work from other assumptions on
the properties of solution near the typhoon eye, is the same as in
\cite{BVDD,BVDD1,BVDD2}, if the first approximation for velocity
and the second approximation for density are made. In this case we
obtain a {\it closed} system for finding an {\it exact} (not only
approximate!) solution. Our system is more less complicated that
one of \cite{BVDD,BVDD1,BVDD2} for second approximation, in
particular cases it can be solved explicitly, however the possible
trajectories are sufficiently realistic. (Note that from numerical
results of \cite{BVDD,BVDD1,BVDD2} one can conclude that the
trajectory depends little on the quadratic terms in the
development of velocity, whereas their presence complicates the
system significantly.)

Moreover, by means of our results we can explain in a sort the
sudden changing of the trajectory direction, a fast decay of
typhoons over dry land, the impossibility of the typhoons
existence in low and high latitudes.

We realize that by means of "toy" models one cannot describe the
complicated typhoon behavior, however, it is interesting that some
important qualitative features can be found.

\section{Simplified model of the typhoon dynamics}

Let ${\bf x}=(x_1,x_2)$ be a point on the Earth surface, $t$ be a
time, $\varphi_0$ be a latitude of some fixed point ${\bf x}_0$,
$\omega = (0,0,\omega_3)^T$ be the angular velocity of the Earth
rotation. Acting in the spirit of \cite{Alishaev} (earlier this
approach was used in \cite{Obukhov} for barotropic atmosphere) we
can get from the standard compressible Navier-Stokes system
\cite{Landau} written for baroclinic rotating atmosphere a
two-dimensional in space system, which as the plane approximation
near ${\bf x}_0$ has the form :
$$
\partial_t {\bf U} +  ({\bf U} \cdot \nabla ){\bf U} +
l T {\bf U} +\varrho^{-1} \nabla p = \mu  \Delta {\bf U} + \lambda
\nabla ( \nabla \cdot {\bf U}),
$$
$$
\partial_t \varrho +  \nabla \cdot ( \varrho {\bf U}) =0,\eqno(NS)
$$
$$
\partial_t S +  ({\bf U}\cdot \nabla S) =0,
$$
where ${\bf U}(t,{\bf x})$ is the velocity vector, $\varrho(t,{\bf
x})>0$ is the density, $S(t,{\bf x})$ is the entropy, $p(t,{\bf
x})$ is the pressure, $ l =2 \omega_3 \sin \varphi_0 $ is the
Coriolis parameter,$ \quad T = \pmatrix{ 0 & -1 \cr 1 & 0 \cr},\,$
$\mu>0$ and $ \lambda $ are the turbulent viscosity coefficients
(may be not only constants).

We supply (NS) by the standard state equation
$$ p=e^S
\varrho^{\gamma},$$ with the adiabatic exponent $\,\gamma>1.$

To obtain this system we can use the unpenetrability conditions on
the Earth surface, the quasi-geostrophic approximation and the
assumption of boundedness of the potential and kinetic energies,
impulse and stream of energy in the air column  (it results the
convergence of all integrals that we need).

Let us denote by $\varrho_\ast, {\bf U}_\ast, p_\ast$ the usual
three-dimensional density, velocity and pressure, respectively,
which are functions of time, the horizontal coordinates,
$x_1,\,x_2,$ and the vertical coordinate, $z.$ Following
\cite{Alishaev}, for arbitrary functions $\phi$ and $f$ we involve
a special notation for averaging over hight, namely, $
\displaystyle\hat\phi =\int_0^\infty \phi\,dz,\quad \bar f
=\frac{1}{\hat\varrho_\ast}\int_0^\infty \varrho_\ast f\,dz.$
Then, $\varrho=\hat\varrho_\ast, \,p=\hat p_\ast,\,{\bf
U}=\bar{\bf U}_\ast.$ Moreover, the usual adiabatic exponent,
$\gamma_\ast=\frac{c_p}{c_v},$ is connected with the
"two-dimensional" adiabatic exponent $\gamma$ as follows:
$\gamma=\displaystyle\frac{2\gamma_\ast-1}{\gamma_\ast}<\gamma_\ast.$
Note that the adiabatic exponent $\gamma$ is later that 2, it is a
very important fact, as we will see below.

Introduce a new variable $\pi=p^{\frac{\gamma-1}{\gamma}}.$  For
the new unknown variables $\pi, {\bf U}, S$ we have the system
$$
\partial_t {\bf U} +  ({\bf U} \cdot \nabla ){\bf U} +
l T {\bf U} + \frac{\gamma}{\gamma-1}\exp{\frac{S}{\gamma}}\nabla
\pi = \mu \Delta {\bf U} + \lambda \nabla ( \nabla \cdot {\bf U}),
$$
$$
\partial_t \pi +  \nabla \pi \cdot  {\bf U} +
(\gamma-1) \nabla  {\bf U}=0, $$
$$
\partial_t S +  ({\bf U}\cdot \nabla S) =
0.$$

Following \cite{BVDD,BVDD1,BVDD2}, we change the coordinate
system, so that the origin of the new system is in the typhoon
eye. Now ${\bf U}= {\bf u}+{\bf V},$ where ${\bf
V}(t)=(V_1(t),V_2(t))$ is a speed of the eye propagation. Thus, we
obtain the new system $$
\partial_t {\bf u} + ({\bf u} \cdot \nabla ){\bf u} +
\dot {\bf V}+ l T ({\bf u}+{\bf V})
+\frac{\gamma}{\gamma-1}\exp{\frac{S}{\gamma}}\nabla \pi = \mu
\Delta {\bf u} + \lambda \nabla ( \nabla \cdot {\bf u}), \eqno
(2.1)
$$
$$
\partial_t \pi +  \nabla \pi\cdot {\bf u} + (\gamma-1)\pi\nabla{\bf u}
= 0, \eqno (2.2) $$
$$
\partial_t S +  ({\bf u}\cdot \nabla S) =0.\eqno(2.3)
$$

Note that in \cite{BVDD,BVDD1,BVDD2} the particular case of (NS)
is considered, namely, $\gamma=2,\,S=const., \lambda=\mu=0.$

Given the vector ${\bf V}$, the trajectory can be found by
integration of the system $$\dot x_1(t)=V_1(t),\qquad \dot
x_2(t)=V_2(t).\eqno(2.4)$$

\section{Solution with linear profile of velocity}

Let us suppose that the velocity vector near the origin has the
form
$${\bf u}(t,{\bf x})=a(t){\bf r}+b(t){\bf r}_\bot,\eqno(3.1)$$ where $${\bf
r}=(x_1,x_2)^T,\, {\bf r}_\bot=(x_2,-x_1)^T.$$
Note that it is possible to construct the solution all over the
plane \cite{RozLANL,RozPas,RozMS}, such that conservation laws
take place, however it is not realistic for our problem, as it
requires the vanishing of density as $|{\bf x}|\to\infty.$ Note
that the velocity with linear profile does not feel the viscosity
term, therefore our result would be the same in the inviscous
model ($\mu=\lambda=0$).

Further, we seek other components of solution to (2.1--2.3) near
the origin in the form
$$\pi(t,{\bf
x})=A(t)x_1^2+B(t)x_1x_2+C(t)x_2^2+M(t)x_1+
N(t)x_2+K(t),\eqno(3.2)$$
$$S(t,{\bf
x})=S_0(t)+\sum\limits_{k_1,k_2=1}^{\infty}\,S_{k_1
k_2}(t)x_1^{k_1} x_2^{k_2}.\eqno(3.3)$$

 From the physical sense $K(t)>0.$ We substitute
(3.1 -- 3.2) in (2.1 -- 2.3) and equal the coefficients at the
same degrees. Firstly, we obtain that $S_{k_1 k_2}(t)\equiv 0,
\,S_0(t)=const.,\, A(t)=C(t),\,B(t)\equiv 0.$ In the center of
typhoon there is a domain of lower pressure, therefore it is
natural to consider $A(t)>0.$

Note that we obtain that the motion near the typhoon center under
our assumptions is "barotropic". However, this barotropicity is
only for the bidimensional  density and pressure, in the usual
sense this does not hold, generally speaking.

Let us introduce a constant $
p_0:=\frac{\gamma}{\gamma-1}\exp{\frac{S_0(0)}{\gamma}}.$

The functions $a(t), b(t), A(t), M(t), N(t), K(t), V_1(t), V_2(t)$
satisfy the following system of ODE: $$\dot A+2\gamma
aA=0,\eqno(3.4)$$ $$\dot a+a^2-b^2+lb+2p_0 A=0,\eqno(3.5)$$
$$\dot b+2ab-la=0,\eqno(3.6)$$ $$\dot
K+2(\gamma-1)aK=0,\eqno(3.7)$$ $$\dot
M+(2\gamma-1)aM-bN=0,\eqno(3.8)$$ $$\dot
N+(2\gamma-1)aN+bM=0,\eqno(3.9)$$ $$\dot V_1-lV_2+p_0
M=0,\eqno(3.10)$$ $$\dot V_2+lV_1+p_0 N=0.\eqno(3.11)$$

 From (3.4) and (3.6) we have
$$b=\frac{l}{2}+C_1|A|^{1/\gamma},\quad  A\ne 0,\eqno(3.12)$$ with a constant
$C_1,$ therefore system (3.4 -- 3.6) can be reduced to  equations
$$\dot A=-2\gamma aA,\eqno(3.4)$$$$\dot
a=-a^2+\frac{l^2}{4}+C_1^2A^{2/\gamma}-2p_0 A.\eqno(3.13)$$

Further, if we know $A(t)$ and $a(t),$ we can find other
functions. Namely, from (3.8), (3.9) we get
$$M(t)=(M^2(0)+N^2(0))^{1/2}\exp\left(-\frac{2\gamma-1}{2}\int\limits
_0^t a(\tau) d\tau\right) \sin(\frac{l}{2}t+C_1\int_0^t
A^{1/\gamma}(\tau)d\tau +C_2),$$
$$N(t)=(M^2(0)+N^2(0))^{1/2}\exp\left(-\frac{2\gamma-1}{2}\int\limits_0^t
a(\tau) d\tau\right) \cos(\frac{l}{2}t+C_1\int_0^t
A^{1/\gamma}(\tau)d\tau +C_2).$$  From (3.4), (3.7) we obtain
$$K(t)=C_3(|A(t)|)^{\frac{\gamma-1}{\gamma}}.$$  Here $C_2, C_3$ are
constants depending only on initial data. However, as follows from
(3.10), (3.11), (2.4) the trajectory does not depend on $K(t).$

Note that in the frame of \cite{BVDD,BVDD1} as the first
approximation on can obtain  system (3.4--3.11) only for $A=0.$ In
\cite{BVDD,BVDD1} it is solved explicitly for $\gamma=2.$

\subsection{Phase plane}

The phase curves of (3.4), (3.13) can be found explicitly. They
satisfy the algebraic equation $$a^2=C_2
A^{\frac{1}{\gamma}}-C_1^2A^{\frac{2}{\gamma}}+\frac{l^2}{4}+
\frac{2p_0}{\gamma-1}A,\eqno(3.14)$$ with a constant $C_4$
depending only on initial data.

We consider below all formally possible values of the constant
$\gamma,$ however, as we have seen, from physical sense
$\gamma<2.$

An elementary  analysis shows that on the phase plane $(A,a)$
there are following equilibria.
\begin{itemize}

\item {If $\,\gamma=2,$ then there exist three equilibria. Namely,
$\,(0, \,\displaystyle \frac{l}{2}),$ a stable node,$\,(0,
\,-\displaystyle \frac{l}{2}),$ unstable node, $\,(\displaystyle
\frac{l^2}{4(2p_0-C_1^2)},\,0),$ a saddle point.}

\item {If $\,\gamma>2,$ then there exist four equilibria. Namely,
$\,(0, \,\displaystyle \frac{l}{2}),$ a stable node, $\,(0,
\,-\displaystyle \frac{l}{2}),$ an unstable node, $\,(A_\pm,\,0),$
where $\,A_\pm$ are roots to the equation $\,\displaystyle
\frac{l^2}{4}+C_1^2 A^{2/\gamma}-2p_0 A=0$ (there are 2 roots of
different signs), they are saddle points.}

\item {If $\,1<\gamma<2$ and $\,f(A_0):=\displaystyle
\frac{l^2}{4}+C_1^2 A_0^{2/\gamma}-2p_0 A_0<0,$ where
$\,A_0=\displaystyle\left(\frac{p_0\gamma}{C_1^2}\right)^
{\frac{\gamma}{2-\gamma}},$ then there exist four equilibria.
Namely, $\,(0, \,\displaystyle \frac{l}{2}),$ a stable node,$\,(0,
\,-\displaystyle \frac{l}{2}),$ an unstable node, $\,(A_\pm,\,0),$
where $A_\pm (A_-<A_+)$ are roots to the equation $\,f(A)=0,$ they
are positive for $l\ne 0$. Moreover, $(A_-,0)$ is a saddle point,
$(A_+,0),$ is a center, if $0<A_-<A_+$.  If $l=0,$ then $\,(0,
\,\displaystyle \pm\frac{l}{2})$ and  $\,(A_-,\,0)$ merge into one
stable-instable equilibrium in the origin.}

 \item{If
$\,1<\gamma<2$ and $\,f(A_0)>0,$ then there exist two equilibria.
Namely, $(0, \,\displaystyle \frac{l}{2}),$ a stable
node,$\,(0,\,-\displaystyle \frac{l}{2}),$ an unstable node.}

\item{If $\,1<\gamma<2$ and $\,f(A_0)=0,$ then there exist three
equilibria. Namely, $\,(0, \,\displaystyle \frac{l}{2}),$ a stable
node,$\,(0, \,-\displaystyle \frac{l}{2}),$ an unstable node,
$\,(A_\pm,\,0),$ where $A_\pm=A_-=A_+,$ a center.}

\end{itemize}

Is is known that typhoons in the phase of maturity behaves as a
stable vortex, moving during a rather long time with a divergency
oscillating about zero and an almost constant vorticity.  We can
see from the phase plane analysis that only in the case $\gamma<2$
there is a possibility of such equilibrium (the center).

Further, it is natural to relate the equilibrium
$(A=0,a=\displaystyle\frac{l}{2})$  to a decaying typhoon. It
follows from (3.12) that $\displaystyle b\to \frac{l}{2}$ as $A\to
0,\,C_1\ne 0.$ However, if $C_1=0$ (where $\displaystyle
b(0)=\frac{l}{2},\,A(0)>0),$ then, on the contrary, there is a
possibility of unrestricted rise of $A(t)$ and $|a(t)|$ (see
below), it also signifies a disappearance of stable structure.

\subsection {Blowup solutions}

From (3.4), (3.14) we get
$$\dot A=\pm
2\gamma A\sqrt{C_4A^{1/\gamma}- C_1^2 A^{2/\gamma} +\frac{2
p_0}{\gamma-1}A+\frac {l^2}{4}}.\eqno(3.15)$$ For $C_1= C_4=0$
equation (3.15) can be integrated explicitly.

To find the equilibrium, we can find zeros of the function
$$
\Phi(A)=C_4A^{1/\gamma}- C_1^2 A^{2/\gamma} +\frac{2 p_0
}{\gamma-1}A+\frac{l^2}{4}.\eqno(3.16)$$ For the case $\gamma\ge
2,\,$ the greatest power in (3.16) is the first. If $\gamma>2$,
the corresponding coefficient is positive, therefore, as follows
from (3.14), $A\to\infty$ as $|a(t)|\to \infty.$ From (3.4),
(3.13)we can see that if  $C_1^2\le \displaystyle
2p_0A^{\frac{2-\gamma}{\gamma}}(0),\,$$
a(0)<-\displaystyle\frac{l}{2},$ then $A(t)$ increases. Really, if
$a(t)<0,\,A(t)>0,$ then $\dot A>0,$ to guarantee $a(t)<0$ we can
use assumptions on $A(0),\,C_1.$

In particular, $\displaystyle C_1^2 A^{\frac{2}{\gamma}-1}-2p_0\le
0,$ if it holds initially. Further, from (3.13) for $l\ne 0$
taking into account this inequality we obtain
$$\dot a\le -a^2+\frac{l^2}{4},\qquad a\le \frac{l}{2}\frac{\tilde
Ce^{lt}+1}{\tilde C e^{lt}-1},\eqno(3.17)
$$ where $\displaystyle \tilde C=\frac{2a(0)+l}{2a(0)-l}<1.$
Therefore $a(t)\to -\infty,$ as $\displaystyle t\to
T=\frac{1}{l}\ln\frac{1}{\tilde C}<\infty.$ It may be interpreted
as a formation of quickly moving narrow vortex (the spout). For
$l=0, $ the analogous result can be easily obtained.

However, if $\gamma\in (1,2),$ then the highest power in (3.16) is
$\displaystyle\frac{2}{\gamma}.$ The coefficient of this term is
non-positive. Therefore  $A(t)$ can tend to $+\infty$ only if
$C_1=0,$ that is in the case $b(0)=\displaystyle\frac{l}{2}.$

Note that in the real atmosphere $l$ depends on the latitude and
only the case $\gamma\in (1,2)$ has a physical sense. As a typhoon
cannot move along a parallel (see below, section 4), then the
blowup phenomenon for a vortex of constant vorticity does not seem
realistic.

\subsection{Why typhoons do not exist in low and high
latitudes?}

It is well known that typhoons never appear lower then $5^o$ and
higher then $30^o$ of latitude, however the mature vortex sometime
goes up to $45^o.$ Let us show that this fact one can explain by
means of our simple model, taking into account only the
relationship between the relative vorticity and the Coriolis
parameter.

It seems naive to explain the phenomenon without temperature
factors, convection and global circulation, however in \cite{Gray}
among (experimentally found) factors, putting the typhoon
development ahead, foremost ones are the initial relative
vorticity and the Coriolis parameter. We stress once more that we
try describe the situation only qualitatively, and we study
conditions of existence of intense vortex, not of its appearance.

Recall that the stable equilibrium (the center) can exist in our
model only if $\gamma<2$ and if the function $f(A)=\displaystyle
\frac{l^2}{4}+C_1^2 A^{\frac{2}{\gamma}}-2p_0 A$ takes a negative
value at some $A>0.$ Let us consider this function as a function
of parameter $l,$ other parameter being fixed. Namely, $$\tilde
f(l)=\displaystyle
\frac{l^2}{4}+\left(b(0)-\frac{l}{2}\right)^2\left(\frac
{A}{A(0)}\right)^{\frac{2}{\gamma}}-2p_0 A.$$ For the sake of
simplicity we assume that we are in the equilibrium point, that is
$A(t)=A(0)=A_+,\,b(t)=b(0):=b_0,$ therefore
$$\tilde f(l)= \displaystyle \frac{l^2}{2}+ b_0 l +b_0^2 -2p_0
A_+.$$ We see that $\tilde f(l)$ can be negative only if
$b_0^2<4p_0A_+,$ moreover, we suppose that $b_0^2>2p_0A_+,$ that
is the vorticity is sufficiently intense. Then $\tilde f(l)<0$
only if $$l\in (l_-,l_+),
\,0<l_-<l_+,\,l_\pm=b_0\pm\sqrt{4p_0A_+-b^2_0}.$$ Thus, we find
the restriction for the Coriolis parameter, necessary for the
existence of stable equilibrium $(A=A_+>0,\,a=0)$ on the phase
plane of system (3.4), (3.13).

\subsection{Why typhoons do not exist over a dry land?}

Let us show that in the frame of our model one can explain the
fact that typhoons do not exist over a dry land (though, no doubt,
the process of evaporation  not taken into account also plays an
important role).

The key point is the significant increasing of the dry friction
when the typhoon goes to the land. Now in the Navier-Stokes system
(NS) in the right hand side of equation for the velocity there
arises the damping term, $\, -\kappa {\bf U},$ where $\,\kappa\,$
is a positive function of coordinates, for the sake of simplicity
we assume that is is a constant. Therefore instead of (3.4),(3.6)
we get
$$\dot a+a^2-b^2+lb+2p_0 A=-\kappa a,\eqno(3.18)$$
$$\dot b+2ab-la=-\kappa b,\eqno(3.19)$$
equations (3.4),(3.7 -- 3.11) do not change.

System (3.4), (3.18), (3.19) is closed, however now it does not
possess any equilibrium such that $A\ne 0,$ therefore we cannot
hope to find a stable domain of low pressure.

\section{Possible trajectories}

Let us analyze trajectories of a stable typhoon, that is we
suppose that $A=A_+,$ where $A_+$ is the greatest zero of the
function $\Phi(A)$ in (3.16), $a=0,\, b=b_0= \displaystyle
\frac{l}{2}+C_1( A_+)^{\frac{1}{\gamma}}.$ From (3.8 -- 3.11),
(2.4) we obtain in this case

if $l\ne b_0,$ then $$x_1(t)=x_1(0)+\frac{V_2(0)}{l}+\frac{p_0
M(0)}{b_0 l} +$$$$\left(\frac{V_1(0)}{l}-\frac{p_0
N(0)}{l(b_0-l)}\right)\sin lt -\left(\frac{V_2(0)}{l}+\frac{p_0
M(0)}{l(b_0-l)}\right)\cos lt+$$ $$ \frac{p_0
N(0)}{b_0(b_0-l)}\sin b_0t+\frac{p_0 M(0)}{b_0(b_0-l)}\cos b_0t,$$

 $$x_2(t)=x_1(0)-\frac{V_1(0)}{l}+\frac{p_0 N(0)}{b_0 l}
+$$$$\left(\frac{V_2(0)}{l}+\frac{p_0 M(0)}{l(b_0-l)}\right)\sin
lt +\left(\frac{V_1(0)}{l}-\frac{p_0 N(0)}{l(b_0-l)}\right)\cos
lt-$$ $$ -\frac{p_0 M(0)}{b_0(b_0-l)}\sin
b_0t+\frac{p_0N(0)}{b_0(b_0-l)}\cos b_0t;$$ if $l= b_0,$ then
$$x_1(t)=x_1(0)+\frac{V_2(0)}{l}+\frac{p_0M(0)}{l^2}
+$$$$\left(\frac{V_1(0)}{l}-\frac{p_0N(0)}{l^2}-\frac{p_0M(0)t}{l}\right)\sin
lt
-\left(\frac{V_2(0)}{l}+\frac{p_0M(0)}{l^2}-\frac{p_0N(0)t}{l}\right)\cos
lt,$$$$x_2(t)=x_2(0)-\frac{V_1(0)}{l}+\frac{p_0N(0)}{l^2}
+$$$$\left(\frac{V_2(0)}{l}+\frac{p_0M(0)}{l^2}-\frac{p_0N(0)t}{l}\right)\sin
lt +\left(\frac{V_1(0)}{l}-\frac{p_0N(0)}{l^2}-\frac{p_0
M(0)t}{l}\right)\cos lt.$$

Thus, we can consider several cases.

I. $\displaystyle |\frac{p_0^2(M^2(0)+N^2(0))}{b_0^2(b_0-l)^2}|<<
\frac{1}{l^2}\left(\left(V_1(0)-\frac{p_0N(0)}{b_0-l}\right)^2+
\left(V_2(0)+\frac{p_0 M(0)}{b_0-l}\right)^2\right).$

It takes place, for example, if $|b_0|>> |l|,$ that is the vortex
is rotating fast. Thus, the trajectory is very close to the
circumference of radius
$$\displaystyle\frac{1}{|l|}\left(\left(V_1(0)-\frac{p_0
N(0)}{b_0-l}\right)^2+ \left(V_2(0)+\frac{p_0
M(0)}{b_0-l}\right)^2\right)^{1/2}\eqno(4.1)$$  with the center in
$$\displaystyle\left(
x_1(0)+\frac{V_2(0)}{l}+\frac{p_0M(0)}{lb_0},\,x_2(0)-\frac{V_1(0)}{l}+
\frac{p_0 N(0)}{lb_0}\right).$$ Figure 1 presents this situation.
The movement is resolute, the typhoon looks like as a single whole
carrying by the main flow.

Note that the decaying typhoon $(a\to \displaystyle \frac {l}{2})$
asymptotically has the analogous trajectory, because $M(t)$ and
$N(t)$ vanish as $t\to\infty.$

II.$\displaystyle
|\frac{p_0^2(M^2(0)+N^2(0))}{b_0^2(b_0-l)^2}|\sim
\frac{1}{l^2}\left(\left(V_1(0)-\frac{p_0 N(0)}{b_0-l}\right)^2+
\left(V_2(0)+\frac{p_0 M(0)}{b_0-l}\right)^2\right).$

In this case
two circular movements superpose. It leads to the appearance of
loops, sudden changing of direction and other complicated
trajectories. Several examples are presented on Figures 2 and 3.

III. $\displaystyle
|\frac{p_0^2(M^2(0)+N^2(0))}{b_0^2(b_0-l)^2}|>>
\frac{1}{l^2}\left(\left(V_1(0)-\frac{p_0N(0)}{b_0-l}\right)^2+
\left(V_2(0)+\frac{p_0 M(0)}{b_0-l}\right)^2\right).$

Here the movement is also almost circular, but its radius is
$\,\displaystyle\frac{(p_0(M^2(0)+N^2(0)))^{1/2}}{|b_0(b_0-l)|}.$

IV. $b_0=l,$ the resonance case. The motion is spiral, moreover,
within some time the vortex can approach to the point with the
coordinates $\left(\displaystyle x_1(0)+\frac{V_2(0)}{l}+\frac{p_0
M(0)}{l^2},\,x_2(0)-\frac{V_1(0)}{l}+\frac{p_0 N(0)}{l^2}\right),
$ and then move away. One of the possible situation is presented
on Figure 4. Stress that this case hardly can  be realized in the
atmosphere, as the value of $l$ changes with the latitude.

In the situations presented on the Figures
$x_1(0)=x_2(0)=1000\,(km),$ $\,p_0=10^5.$

\begin {figure}
\scalebox {1.35}{\includegraphics [1,150]{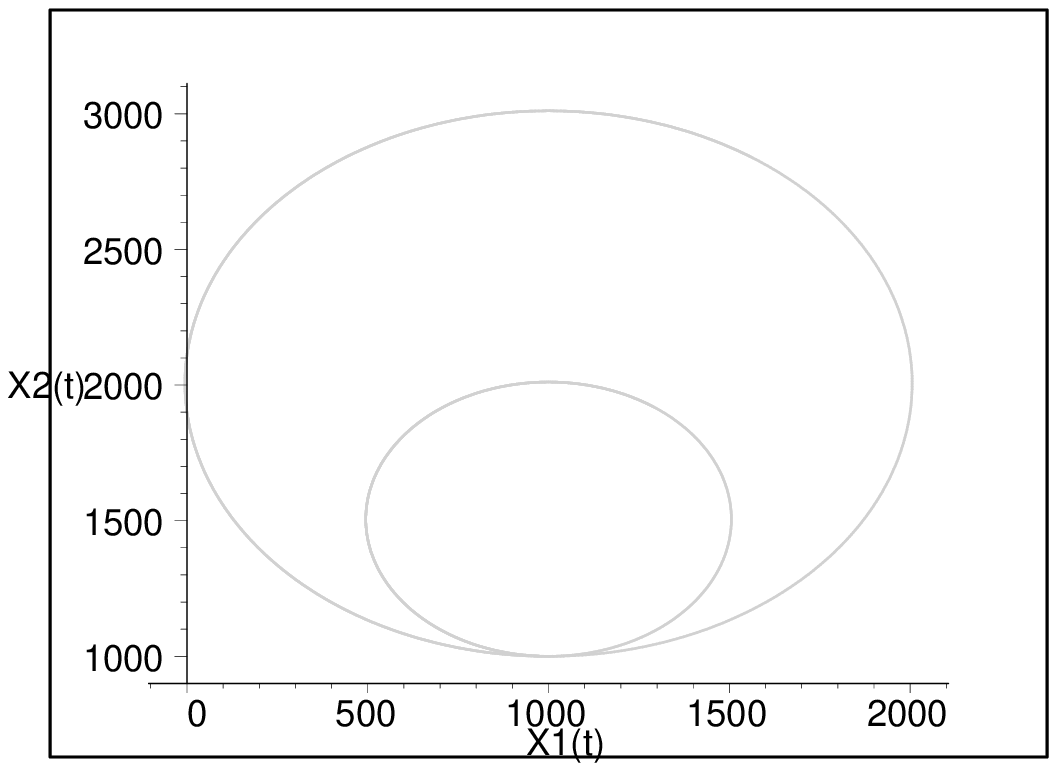}} \caption
{Circular trajectory (quickly rotating typhoon)}
\end {figure}

{\sc Figure 1.} A circular trajectory. Here $l=10^{-5}\, (s^{-1}),
\,$$ b_0=2l,\,$$M(0)=10^{-14}\,, $ $N(0)=0,\, V_1(0)=-5\, (mps),$$
\,V_2(0)=0,$ (the minor circle) $V_1(0)=-10 \,(mps), \,V_2(0)=0$
(the larger circle).

\begin {figure}
\scalebox {1.35}{\includegraphics [1,150]{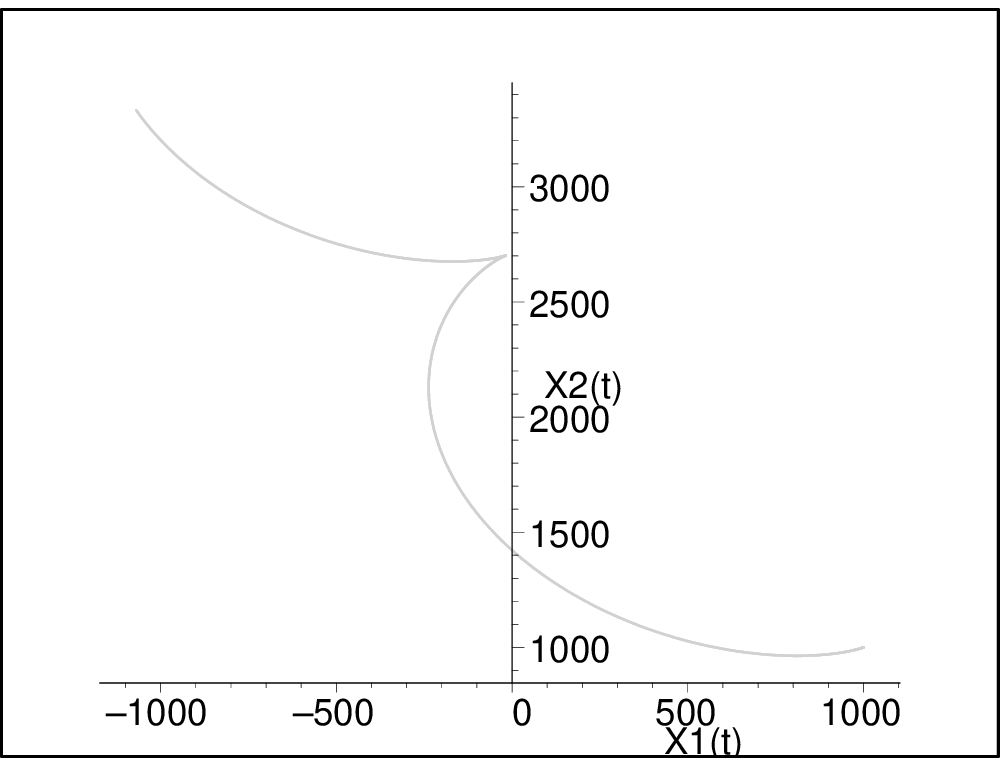}}
\caption {Slowly rotating typhoon}
\end {figure}
\begin {figure}
\scalebox {1.35}{\includegraphics [1,150]{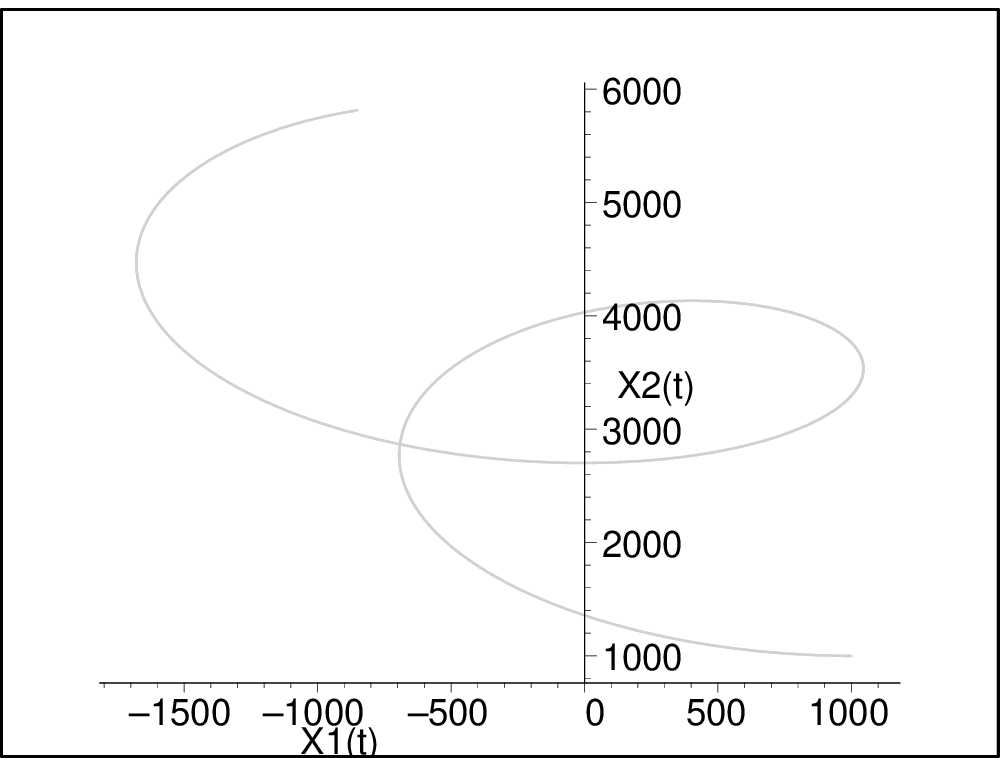}}
\caption {Slowly rotating typhoon}
\end {figure}

{\sc Figures 2 and 3.} A slowly rotating typhoon (loops and change
of direction). Here $l=8\cdot 10^{-5} \, (s^{-1}), \,
b_0=l/100,\,$ $M(0)=2\cdot 10^{-14}\,, \,$$ N(0)=0,\,$$
\,V_1(0)=V_2(0)=0,$ (Fig.2) $V_1(0)=-10\, (mps), \,V_2(0)=0$
(Fig.3).

\begin {figure}
\scalebox {1.35}{\includegraphics [1,150]{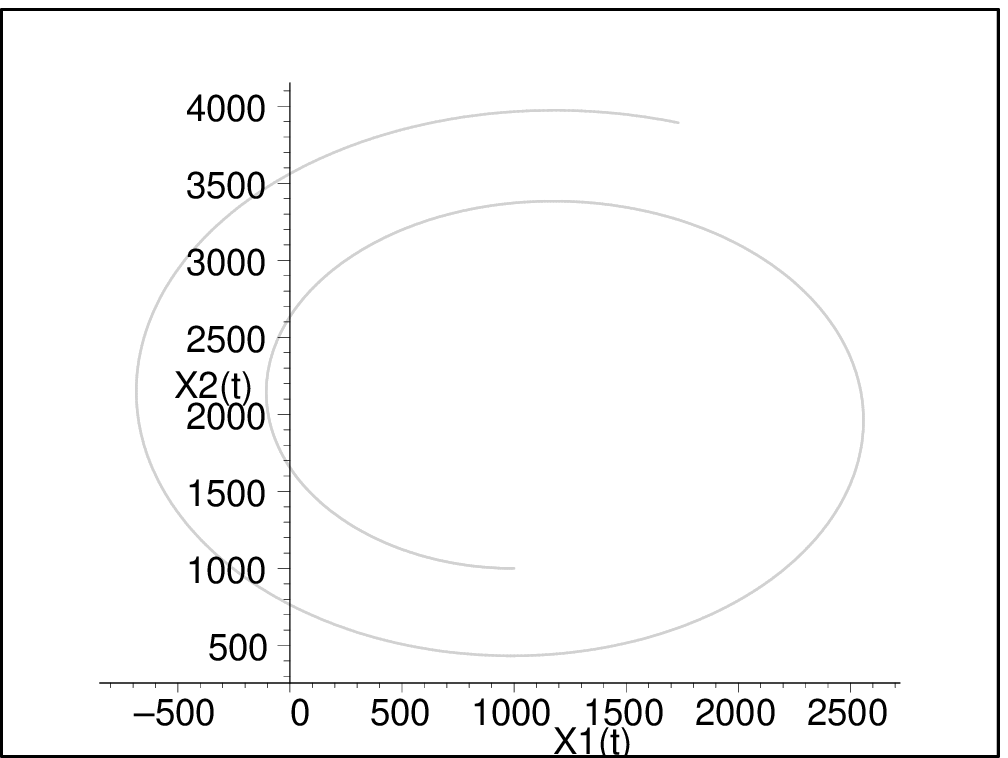}}
\caption {Resonance case}
\end {figure}

{\sc Figure 4.} Resonance case. Here $l=b_0=10^{-5}\, s^{-1}, \,
M(0)=10^{-13}\,, \,$$ N(0)=0,\,$$ V_1(0)=-10\, (mps), \,V_2(0)=0.$

\begin {figure}
\scalebox {1.35}{\includegraphics [1,150]{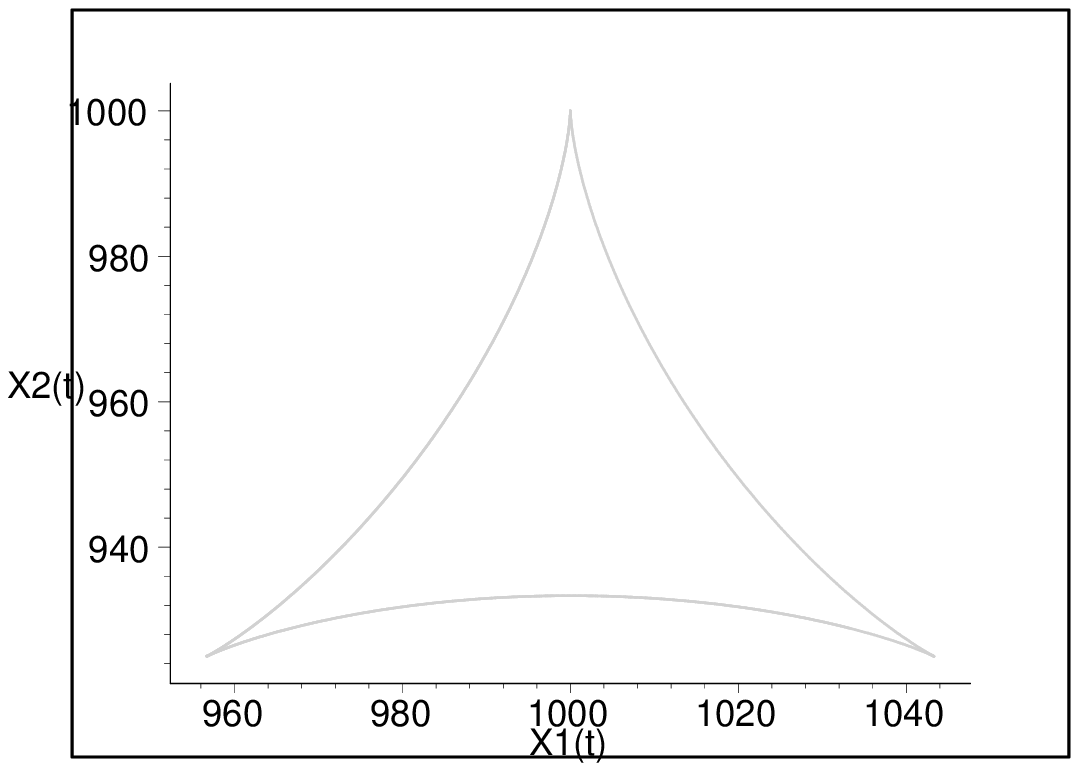}} \caption {
"Making time" typhoon}
\end {figure}

\begin {figure}
\scalebox {1.35}{\includegraphics [1,150]{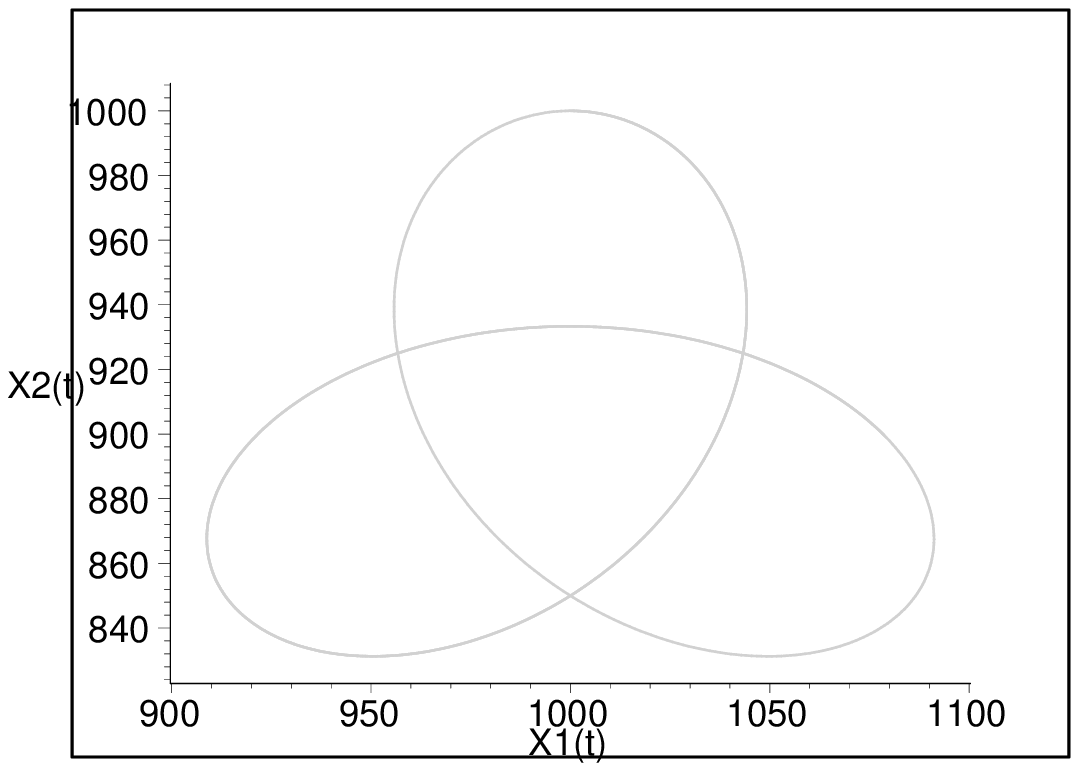}} \caption {
"Making time" typhoon}
\end {figure}

{\sc Figures 5 and 6.} A "making time" typhoon. The examples of
exotic trajectories are presented. Here $l=2\cdot 10^{-5}\,
s^{-1}, \,b_0=-\displaystyle\frac{l}{2}, $$\,M(0)=10^{-13}\,, \,
N(0)=0,\,V_1(0)=V_2(0)=0$ (Fig.5), $ V_1(0)=1 \,(mps), \,V_2(0)=0$
(Fig.6).

Let us stress, that as usual, except several cases, in the real
atmosphere the typhoon doesn't pass throughout all the trajectory
before its decay, and one can speak about a part of this
trajectory. Moreover, because the eye is not exactly in the
equilibrium point, its trajectory oscillates near the trajectory
presented in the picture.

Note that the rather usual behavior of intense typhoons is the
motion first along the almost circular trajectory, becoming more
curved as $l$ rises, and then along the almost strait trajectory.
This can be explained as follows: if we are in the situation of
quickly rotating typhoon (case I according to the classification
of this section,$\,b_0>>l$), then firstly the difference $b_0-l$
is rather large, and the decisive influence to the value of radius
of trajectory has the Coriolis parameter itself (see formula
(4.1)). However, as the Coriolis parameter raises, this difference
becomes small, it implies the augmentation of radius, thus the
trajectory can look like a strait line. We can give examples of
such typhoons: Isaac,  category: 4 , 21 Sep -- 01 Oct, 2000,
Atlantic; Mitag,  category: 5, 26 Feb -- 08 Mar, 2002, Western
Pacific; Pongsona,  category: 4, 2--11 Dec, 2002, Western Pacific;
Fabian,  category: 4, 27 Aug -- 08 Sep, 2003, Atlantic. (Recall
that the number of category increases with the intensity of
typhoon, the greatest is 5th, where the maximum wind amount to $70
mps$ and higher.)

Typhoons, which trajectories loop, as usual (but not necessarily),
are not very intense. Examples are the following: Fung Wong,
category: 1, 20--27 July, 2002, Western Pacific; Tropical
depression 28W, 18 --24 July, 2001, Western Pacific; Kyle,
category: 1, 30 Sep -- 12 Oct, 2002, Atlantic (however, Danas,
category: 4, 3--12 Sep, 2001, Atlantic; Saomai,  category: 5,
3--16 Sep, 2000,Western Pacific). They can be related to  class
II, on Figs.2 and 3  such trajectory is obtained for "slow"
typhoons, but one can get trajectories with loops and for intense
typhoons, too.

 At last it is possible that the typhoon
passes to other equilibrium. It is known that that it can decay
and regenerate several times. However it signifies that the
trajectory of its motion modifies according to a new regime. Thus,
as a  rough approximation, the trajectory will be glued from
several standard parts.

\subsubsection*{Forecasting possibilities} As we can see, in the
frame of our model the trajectory behavior completely depends on
initial data. The finding of correct initial data is a separate
difficult problem, not only mathematical, but physical and
statistical. The procedure of self-adapting (described in the
introduction) may be of great use for this purpose.

\subsubsection*{Acknowledgements} The work was partially
supported by the Russian Foundation of Basic Research Award
no.03-02-16263 and the Leading Scientific Schools Project
no.1464.2003.1.

The author is grateful to H.Fujita Yashima for attracting the
attention to the subject and discussion, and to E.R.Rozendorn and
V.A.Gordin for the interest to the work and useful commentaries.

\end{document}